\documentclass[12pt, a4paper]{article}
\usepackage{graphicx}
\usepackage{psfrag}
\usepackage{amsmath}
\usepackage{mathtext}
\usepackage{amssymb}
\usepackage{feynmf}
\newcommand{\beqn}{\begin{eqnarray}}
\newcommand{\eeqn}{\end{eqnarray}}
\newcommand{\beqnn}{\begin{eqnarray*}}
\newcommand{\eeqnn}{\end{eqnarray*}}

\newcommand{\beq}{\begin{equation}}
\newcommand{\eeq}{\end{equation}}
\newcommand{\benq}{\begin{equation}}
\newcommand{\eenq}{\end{equation}}
\newcommand{\Tr}{{T\!r}}
\newcommand{\off}{o\!f\!f}

\def\M{{\cal M}}
\def\U{{U_{x,\mu}}}
\def\Ud{{U^\dagger_{x,\mu}}}
\newcommand{\ds}{\displaystyle}

\newcommand{\calL}{{\cal L}}
\newcommand{\abstracts}[1]{{\centering{\begin{minipage}{13.0truecm}
 \normalsize\baselineskip=15pt \centerline{\footnotesize
 ABSTRACT}\vspace*{0.3cm} \parindent=20pt #1 \end{minipage}}\par}}

\begin{document}
~ \vspace{-1cm}
\begin{flushright}
{\large ITEP-LAT/2005-1}

\vspace{0.2cm}

\vspace{0.3cm}
{\sl \today}
\end{flushright}

\begin{center}
{\baselineskip=24pt {\Large \bf Ultraviolet Behavior of the Gluon
Propagator in the Maximal Abelian Gauge}\\

\vspace{1cm}

{\large
S.~M.~Morozov$^{a}$, R.~N.~Rogalyov$^{b}$
}
}
\vspace{.5cm}
{\baselineskip=16pt
{ \it

$^{\rm a}$ Institute of Theoretical and  Experimental Physics,
Moscow, 117259, Russia\\
$^{\rm b}$ Institute for High Energy Physics, Protvino, Moscow region,
142281, Russia\\
}
}
\end{center}

\vspace{5mm}

\abstracts{The ultraviolet asymptotic behavior of the 
gluon propagator is evaluated in the maximal Abelian gauge in 
the $SU(2)$ gauge theory on the basis of the renormalization-group
improved perturbation theory at the one-loop level.
Square-root singularities obtained in the Euclidean domain
are attributed to artifacts of the one-loop approximation
in the maximal Abelian gauge and the standard normalization 
condition for the propagator used in our study.
It is argued that this gauge is essentially nonperturbative.}

\section{Introduction}

Maximal Abelian Gauge (MAG) \cite{tHooft} 
offers one of the best instruments for theoretical studies
of dynamics of the gauge fields in QCD at a large scale 
($\sim 1$~fm). In this gauge, the boson fields can 
be naturally divided into the "Abelian" and "non-Abelian"
(or, in other words, "diagonal" and "off-diagonal") components.
Assuming that the "diagonal" components give a leading
contribution to the path integral,\footnote{This assumption is
referred to as the hypothesis of Abelian dominance \cite{Poly}.} 
the Yang--Mills theory on a lattice can be reduced to the
dual Abelian Higgs model. In so doing, confinement of quarks
and gluons comes about from monopole condensation in the dual
Abelian Higgs model. 

Therefore, propagators of "diagonal" and "off-diagonal" 
gluons are of particular interest because, from a naive
point of view, their ratio indicates whether the
Abelian dominance hypothesis is true or not.
Recently \cite{Morozov}, the behavior of the
"diagonal" and "off-diagonal" gluon propagators 
was estimated numerically in the framework of lattice gauge theory.
At small momenta it was found that the propagator of the "Abelian"
field components dominates.
In this work, we study asymptotic behavior of the
propagators of the "diagonal" and "off-diagonal" gluon
at large momenta in the framework of perturbation theory
in the continuum field theory\footnote{We work in the Euclidean space,
throughout this paper, $p^2 = p_0^2 + p_1^2 + p_2^2 + p_3^2$, that is,
Euclidean momenta are positive.}. It should be noted that the
interest in this gauge was quickened by the study \cite{Min:1985}, where it was 
demonstrated that the $SU(2)$ Yang-Mills theory can be renormalized in this 
gauge.

\section{Gauge fixing}

In practical calculations in a lattice Yang--Mills theory, 
the maximal Abelian gauge is fixed by finding a stationary 
point of the functional
\beq\label{eq:GaugFunctl}
\Phi= \sum_{x,\mu} \Tr \left( \U \sigma_3 \Ud \sigma_3 \right) 
\eeq
with respect to the gauge transformations of the type
\beq\label{eq:GaugeTransf}
\Lambda: \ \ \ \U \to \Lambda_x^\dagger \U \Lambda_{x+\hat\mu},
\eeq
where $\U \in SU(2)$ are the link variables and $\Lambda_x$ 
is a $SU(2)$-valued function defined at sites of a lattice.
In other words, the gauge condition is fixed by the relations
\beq
\frac{\delta \Phi}{\delta \Lambda} \;=\; 0.
\eeq
At infinitesimal gauge transformations, 
\beq
\Lambda=\exp (-i\omega_a \sigma_a / 2 )=\cos{|\vec \omega | \over 2}
-i {\omega_a \sigma_a \over |\vec \omega |} \sin {|\vec \omega| \over 2},
\eeq
the variation of the functional (\ref{eq:GaugFunctl}) takes the form
\beqnn
\delta\Phi &=& \Tr \left[ \U (\sigma_3 + \M_1(x+\hat\mu) + \M_2(x+\hat\mu)) \Ud
(\sigma_3 + \M_1(x) + \M_2(x)) \right] \\ \nonumber
&& - \Tr [\U \sigma_3 \Ud \sigma_3],
\eeqnn
where
\beq
\M_1(x)=-\epsilon^{3ab} \omega_a(x) \sigma_b, \ \ \ \ \ \ 
\M_2(x)={1\over 2} \left( \delta^{ac}\delta^{b3} - \delta^{ab}\delta^{c3}\right)
\omega_a \omega_b \sigma_c.
\eeq
Using the decomposition
\begin{displaymath}
\U \! = \! u_0+ i \sum_{n=1}^3 u_n\sigma_n \!=\! \left(
\begin{array}{cc}
u_0+ i u_3   &  u_2 + i u_1 \\
-u_2 + i u_1 &  u_0 - i u_3
\end{array}
\right), 
\end{displaymath}
we define the vector potentials $A^a_\mu$  
by the formula 
$\ds u^a_\mu = - \frac{ga}{2} A^a_\mu$ ($a=1,2,3$).
In this case, the gauge condition has the form
\beqn\label{eq:MAGCondLatt}
a \nabla_\mu^B ((u_\mu^3 - iu_\mu^0) (u_\mu^1-iu^2_\mu)) + 2 u_\mu^3 (u_\mu^1-iu^2_\mu) &=& 0, \\ \nonumber
a \nabla_\mu^B ((u_\mu^3 + iu_\mu^0) (u_\mu^1+iu_\mu^2)) + 2 u_\mu^3 (u_\mu^1+iu^2_\mu) &=& 0,
\eeqn
where $\ds \nabla_\mu^B$ is the operator of backward derivative
on a lattice: $\ds \nabla_\mu^B u_i^\nu(x) = {u_i^\nu(x) - u_i^\nu(x-\hat \mu) \over a}$. 
Having regard to the relation $u^0_\mu=1+O(g^2a^2)$, we find that, in 
the leading order in the lattice spacing, the gauge condition
(\ref{eq:MAGCondLatt}) has the form
\beqn\label{eq:MAGCondCont}
&& (\partial_\mu + ig A^3_\mu) (A^1_\mu + iA^2_\mu)=0.\\
&& (\partial_\mu - ig A^3_\mu) (A^1_\mu - iA^2_\mu)=0.
\eeqn
Here and below, this naive limit of the above gauge conditions is named 
maximal Abelian gauge in the continuum theory. 
The continuum theory in this gauge can conveniently be 
quantized in the BRST formalism. To do this, we should introduce
the gauge-fixing terms as follows:
\beq
\Delta {\cal L}_{GF} = - {1\over 2\alpha}\; 
\left| (\partial_\mu + ig A^3_\mu) (A^1_\mu + iA^2_\mu) \right|^2
- {1\over 2 \hat \beta}\; \left( \partial_\mu A^3_\mu \right)^2,
\eeq
where $\alpha$ and $\hat \beta$ are the gauge parameters.
The former term is responsible for the gauge condition (\ref{eq:MAGCondCont}),
the latter term is needed to fix the remaining $U(1)$ gauge arbitrariness.
The case $\alpha= \hat \beta = 1$ corresponds to the case $\xi = 1$ from
article \cite{Suzuki}.

%
%

The BRST invariant Lagrangian of the Yang--Mills field in the maximal
Abelian gauge has the form

\beqnn
	{\cal L} &=& \frac{1}{4}f_{\mu\nu}f_{\mu\nu} 
	+\frac{1}{2\hat{\beta}}(\partial_\mu a_\mu)^2 
	+\frac{1}{4}\tilde{F}_{\mu\nu}^a\tilde{F}_{\mu\nu}^a 
	+ \frac{1}{2\alpha}(\partial_\mu A_\mu^a)^2 \\
	&-& i\kappa\bar{C}^3\partial^2C^3 - i\bar{C}^a\partial^2C^a 
	- \frac{\alpha}{4}g^2\varepsilon^{ab}\varepsilon^{cd}\bar{C}^a\bar{C}^bC^cC^d \\
	&+&\frac{1}{2}g\epsilon^{ab}
	\big(a_\mu A_\nu^b\partial_\nu A_\mu^a - a_\mu A_\nu^a\partial_\nu A_\mu^b
	+ A_\mu^a A_\nu^b\partial_\mu a_\nu - a_\mu A_\nu^b \partial_\mu A_\nu^a \\
	&\ & + a_\mu A_\nu^a \partial_\mu A_\nu^b - A_\nu^a A_\mu^b \partial_\mu a_\nu
	- \frac{1}{\alpha}a_\mu A_\mu^a \partial_\nu A_\nu^b
	+ \frac{1}{\alpha} a_\mu A_\mu^b \partial_\nu A_\nu^a\big) \\
	&-& ig\varepsilon^{ab}a_\mu(C^b\partial_\mu\bar{C}^a - \bar{C}^a\partial_\mu C^b)
	- ig\kappa\varepsilon^{ab}\bar{C}^3\partial_\mu(A_\mu^aC^b) \\
	&+&\frac{1}{4}g^2\delta^{ab}\left(2\delta_{\mu\nu}\delta_{\rho\sigma}
	- (1-\frac{1}{\alpha})(\delta_{\mu\rho}\delta_{\nu\sigma}+
	\delta_{\mu\sigma}\delta_{\nu\rho})\right)
	A_\mu^a A_\nu^b a_\rho a_\sigma \\
	&+&\frac{1}{4}g^2(\delta_{\mu\rho}\delta_{\nu\sigma}-
	\delta_{\nu\rho}\delta_{\mu\sigma})
	A_\mu^aA_\nu^bA_\rho^aA_\sigma^b \\
	&+& ig^2\delta^{ab}\delta_{\mu\nu}\bar{C}^aC^b a_\mu a_\nu 
	+ i g^2(\varepsilon^{ad}\varepsilon^{cb}+\varepsilon^{ac}\varepsilon^{db})
		\bar{C}^aC^bA_\mu^cA_\mu^d,
\eeqnn
where the gluon field $A_\mu^A$ is divided into the "Abelian" and 
"non-Abelian" components as follows:
\beqnn\label{eq:Decomposition}
A_\mu^A T^A = A_\mu^a T^a + a_\mu T^3, 
\eeqnn
where $T^A$ are the $SU(2)$ generators; indices denoted by 
capital letters are assigned one of the integers $1,2,3$
and indices denoted by small letters---$1,2$. 
The gluon fields with the indices denoted by small letters
are named "off-diagonal" and the Abelian component  $a_\mu$ defined by the 
equation (\ref{eq:Decomposition}) is named the field of the "diagonal"
gluon. Let us also introduce the notation: 
\beqnn
\tilde{F}_{\mu\nu}^a &=& \partial_\mu A_\nu^a - \partial_\nu A_\mu^a,  \\
f_{\mu\nu} &=& \partial_\mu a_\nu - \partial_\nu a_\mu,                
\eeqnn
A detailed discussion of the parameter $\kappa$ can be found in \cite{Shinohara:2001cw},
here we set $\kappa = 1$.

\section{Feynman rules}

The expressions for the gluon and ghost propagators
and three- and four-particle vertices are readily obtained by
standard techniques. The results are shown in the table 

\begin{center}
\begin{tabular}{|c|c|c|}
\hline
\multicolumn{2}{|c|}{Propagators} \\
\hline
$ \langle a_\mu(p) a_\nu(-p)\rangle $ &
$ \frac{1}{p^2} \left(\delta_{\mu\nu}-(1-\hat{\beta})\frac{p_\mu p_\nu}{p^2}\right) $\\
\hline
$ \langle A_\mu^a(p) A_\nu^b(-p)\rangle $ &
$ \frac{\delta^{ab}}{p^2} 
\left(\delta_{\mu\nu}-(1-\alpha)\frac{p_\mu p_\nu}{p^2}\right) $ \\
\hline
$ \langle \bar{C}^3(p) C^3(-p) \rangle $ &
$ -\frac{i}{p^2} $ \\
\hline
$ \langle \bar{C}^a(p) C^b(-p) \rangle $ &
$ -i \frac{\delta^{ab}}{p^2} $ \\
\hline
\multicolumn{2}{|c|}{Three-particle vertices} \\
\hline
$ \langle a_\mu(p)A_\rho^a(q)A_\sigma^b(r)\rangle $ & 
$ - i g\varepsilon^{ab}\left( \left(q-r\right)_\mu \delta_{\rho\sigma}
+ \left(r-p+\frac{q}{\alpha}\right)_\rho \delta_{\sigma\mu} \right. $ \\
& $ \left. + \left(r-q-\frac{r}{\alpha}\right)_\sigma \delta_{\mu\rho}\right) $ \\
\hline
$ \langle\bar{C}^a(p)C^b(q)a_\mu\rangle $ & 
$ -g(p-q)_\mu\varepsilon^{ab} $ \\
\hline
$ \langle\bar{C}^3(p)C^b(q)A_\mu^a\rangle $ &
$ -\kappa g\varepsilon^{ab}p_\mu $ \\
\hline
\multicolumn{2}{|c|}{Four-particle vertices} \\
\hline
$ \langle a_\mu a_\nu A_\rho^a A_\sigma^b\rangle $ &
$ - g^2\delta^{ab}\left(2\delta_{\mu\nu}\delta_{\rho\sigma} 
- (1-\frac{1}{\alpha})(\delta_{\mu\rho}\delta_{\nu\sigma}
+ \delta_{\mu\sigma}\delta_{\nu\rho} \right) $ \\
\hline
$ \langle A_\mu^a A_\nu^b A_\rho^c A_\sigma^d\rangle $ &
$ -g^2\left(\varepsilon^{ab}\varepsilon^{cd}I_{\mu\nu,\rho\sigma}+
\varepsilon^{ac}\varepsilon^{bd}I_{\mu\rho,\nu\sigma}+
\varepsilon^{ad}\varepsilon^{bc}I_{\mu\sigma,\nu\rho} \right) $\\
& $	I_{\mu\nu,\rho\sigma} = \left(
\delta_{\mu\rho}\delta_{\nu\sigma}-\delta_{\mu\sigma}\delta_{\nu\rho}\right) $ \\
\hline
$ \langle\bar{C}^a C^b a_\mu a_\nu \rangle $ & 
$ - i 2g^2\delta^{ab}\delta_{\mu\nu} $ \\
\hline
$ \langle \bar{C}^a C^b A_\mu^c A_\nu^d \rangle $ &
$ -i g^2\delta_{\mu\nu} \left(\varepsilon^{ad}\varepsilon^{cb}
+\varepsilon^{ac}\varepsilon^{db}\right) $ \\
\hline
$ \langle \bar{C}^a \bar{C}^b C^c C^d\rangle $ &
$ g^2\alpha\varepsilon^{ab}\varepsilon^{cd} $ \\
\hline
\end{tabular}
\end{center}

It should be noted that the "diagonal" ghost $C^3$ 
interacts with no field, whereas the "diagonal" anti-ghost $\bar{C}^3$ 
interacts with some fields. Therefore, the "diagonal"
ghosts should be disregarded in the loop expansion of
perturbation theory.

\section{Gluon Propagator to One-Loop}
The diagrams contributing to the polarization operator of the
"diagonal" gluon are shown in Fig.~(\ref{fig:diag_conf}).

\begin{fmffile}{diag}
\begin{figure}[h!]
\begin{center}
\begin{tabular}{cccc}
	$\Pi_{\mu\nu}^{33}(p)$ =
	& \parbox{20mm}{\begin{fmfgraph}(20,20)
		\fmfleft{i}
		\fmfright{o}
		\fmf{wiggly}{i,v1}
		\fmffreeze
		\fmfforce{0.2w, 0.5h}{v1}
		\fmfforce{0.8w, 0.5h}{v2}
		\fmf{ghost,left,tag=1}{v1,v2}
		\fmf{ghost,left,tag=2}{v2,v1}
		\fmfdot{v1}
		\fmfdot{v2}
		\fmf{wiggly}{v2,o}
	\end{fmfgraph}}
	& + & \parbox{20mm}{\begin{fmfgraph}(20,20)
		\fmfleft{i}
		\fmfright{o}
		\fmf{wiggly}{i,v1}
		\fmffreeze
		\fmfforce{0.2w, 0.5h}{v1}
		\fmfforce{0.8w, 0.5h}{v2}
		\fmfdot{v1}
		\fmfdot{v2}
		\fmf{gluon,left,tag=1}{v1,v2}
		\fmf{gluon,left,tag=2}{v2,v1}
		\fmf{wiggly}{v2,o}
	\end{fmfgraph}} \\ 
	+ & 
	\parbox{20mm}{\begin{fmfgraph}(20,20)
		\fmfleft{i}
		\fmfright{o}
		\fmf{wiggly}{i,v1}
		\fmf{gluon}{v1,v1}
		\fmfdot{v1}
		\fmf{wiggly}{v1,o}
	\end{fmfgraph}}
	& + & \parbox{20mm}{\begin{fmfgraph}(20,20)
		\fmfleft{i}
		\fmfright{o}
		\fmf{wiggly}{i,v1}
		\fmf{ghost}{v1,v1}
		\fmfdot{v1}
		\fmf{wiggly}{v1,o}
	\end{fmfgraph}}	 
\end{tabular}
\end{center}
\caption{Polarization operator of the "diagonal" gluon.}
\label{fig:diag_conf}
\end{figure}
\end{fmffile}

Our computations are performed using the dimensional regularization
techniques with the space-time dimension $D = 4-2\epsilon$; in so doing,
the contribution of the tadpole diagrams vanishes.

The contribution of the non-vanishing diagrams to the gluon polarization 
operator has the form  
\beqn
\label{eq:diag_pol}
\Pi_{\mu\nu}^{33} &=& \frac{g^2}{16\pi^2} (\delta_{\mu\nu}p^2-p_\mu p_\nu) \left(
\frac{22}{3} L + \left(\frac{205}{18}+3\alpha+\frac{\alpha^2}{2}\right)\right),
\eeqn
where 
\beq
L={1\over \epsilon} - \gamma_E + \ln\left({4\pi\mu^2\over p^2}\right)
\eeq
and $\gamma_E$ is the Euler constant. 


Performing analogous computations for the "off-diagonal" gluon
(see diagrams in Fig.~(\ref{fig:offdiag_cont})), we arrive at 

\begin{fmffile}{offdiag}
\begin{figure}[h!]
\begin{center}
\begin{tabular}{cccc}
	$\Pi_{\mu\nu}^{ab}(p)$ = 
	& \parbox{20mm}{\begin{fmfgraph}(20,20)
		\fmfleft{i}
		\fmfright{o}
		\fmf{gluon}{i,v1}
		\fmffreeze
		\fmfforce{0.2w, 0.5h}{v1}
		\fmfforce{0.8w, 0.5h}{v2}
		\fmf{wiggly,left,tag=1}{v1,v2}
		\fmf{gluon,left,tag=2}{v2,v1}
		\fmfdot{v1}
		\fmfdot{v2}
		\fmf{gluon}{v2,o}
	\end{fmfgraph}}
	& + &
	\parbox{20mm}{\begin{fmfgraph}(20,20)
		\fmfleft{i}
		\fmfright{o}
		\fmf{gluon}{i,v1}
		\fmf{wiggly}{v1,v1}
		\fmfdot{v1}
		\fmf{gluon}{v1,o}
	\end{fmfgraph}} \\
	+ & \parbox{20mm}{\begin{fmfgraph}(20,20)
		\fmfleft{i}
		\fmfright{o}
		\fmf{gluon}{i,v1}
		\fmf{gluon}{v1,v1}
		\fmfdot{v1}
		\fmf{gluon}{v1,o}
	\end{fmfgraph}}
	& + & 
	\parbox{20mm}{\begin{fmfgraph}(20,20)
		\fmfleft{i}
		\fmfright{o}
		\fmf{gluon}{i,v1}
		\fmf{ghost}{v1,v1}
		\fmfdot{v1}
		\fmf{gluon}{v1,o}
	\end{fmfgraph}}	 
\end{tabular}
\end{center}
\caption{Polarization operator of the "off-diagonal" gluon.}
\label{fig:offdiag_cont}
\end{figure}
\end{fmffile}

\beqn
\label{eq:offdiag_pol}
\Pi_{\mu\nu}^{ab}(p) = -\;\delta^{ab}\frac{g^2}{16\pi^2} 
\left( (\delta_{\mu\nu} p^2 - p_\mu p_\nu) T^{\off}(p^2) + 
p_\mu p_\nu L^{\off}(p^2)\right),
\eeqn
where
\beqn
T^{\off}(p^2) &=&
\left(
	-\frac{17}{6}+\beta+\frac{\alpha}{2}
\right) L\,+\,
\left(
	-\frac{43}{18}+\frac{\beta}{2}-\frac{\alpha\beta}{2}-\frac{\alpha}{2}
\right),\\ \nonumber
L^{\off}(p^2) &=&
\left(
	-\frac{1}{2} + \frac{\beta}{\alpha} - \frac{3}{2\alpha} - \frac{3}{\alpha^2}
\right) L\,+\, \frac{1}{2\alpha}
\left(
	-7 + 3\beta - \frac{3\beta}{\alpha} - \frac{5}{\alpha}
\right).
\eeqn


The expressions (\ref{eq:diag_pol}) and (\ref{eq:offdiag_pol})
involve both divergent and finite parts (the divergent part was computed
in \cite{Shinohara:2001cw}). The finite part of 
the longitudinal component of the
"off-diagonal" propagator (\ref{eq:offdiag_pol}) involves terms
proportional to $\frac{1}{\alpha}$ and $\frac{1}{\alpha^2}$, 
which are singular when $\alpha \rightarrow 0$. However, 
they give no contribution to the expressions for physical quantities.
Their presence in the expression for the propagator is due to
the fact that the interaction vertex in this gauge is proportional to
$\frac{1}{\alpha}$, where $\alpha$ is the gauge parameter that appears 
in the expression for the unperturbed propagator. 
For this reason, $\alpha \rightarrow 0$ corresponds to 
very dangerous limit and thus it is safe in the perturbation expansion
to assume a nonzero value of $\alpha$. 


In the $\bar {MS}$ subtraction scheme, the counterterms 
added to the Lagrangian used to compensate for the divergent terms 
are as follows:
\beqn
\Delta{\cal L}_{counter} &=& {(Z_A^{off})^2 -1 \over 4}\; (\partial_\mu A^a_\nu - \partial_\nu A^a_\mu)^2 + 
{X^{off}\over 2} (\partial_\mu A^a_\mu)^2 + \\ \nonumber
&& + \, {(Z_A^{(3)})^2 -1 \over 4} \; (\partial_\mu a_\nu - \partial_\nu a_\mu)^2 +
{X^{(3)}\over 2} (\partial_\mu a_\mu)^2, \\ \nonumber
\eeqn
where 
\beqn
Z^{(3)}_{A} &= & 1 + \;{11 \over 3}\;{g^2\over 16\pi^2 \epsilon'}, \\ \nonumber
Z^{off}_{A} &= & 1 + {g^2\over 16\pi^2 \epsilon'}\;{1\over 2}\;\left({17 \over 6} - \beta - {\alpha \over 2} \right), \\ \nonumber
X^{(3)} &= & 0, \\ \nonumber
X^{off} &= & {g^2\over 16\pi^2 \epsilon'}\; 
\left(\, {1 \over 2}\,-\,{\beta \over\alpha }\,+\, {3 \over \alpha^2} \,+\, {3 \over 2\alpha} \right),\\ \nonumber
\eeqn
and the modified parameter $\epsilon'$ is defined by the formula
\beqnn
{1\over \epsilon'} = {1\over \epsilon} - \gamma_E + \ln (4\pi).
\eeqnn

\section{The renormalization group equation}

\subsection{The case of Lorentz gauge}

First we consider the renormalization group equation in the 
Lorentz gauge in order to demonstrate some features 
of its solution in the maximal Abelian gauge. 
The Lorentz gauge can be specified by the gauge fixing term
in the Lagrangian as follows:
\beq
\Delta \calL_{GF} = \frac{1}{2\alpha} (\partial_\mu A_\mu^A)^2.
\eeq

The gluon propagator in this gauge is parametrized by only one
scalar function
\beq
\label{eq:PropLor}
    G_{\mu\nu}^{AB}(p) = \frac{\delta^{AB}}{p^2} \;\left( 
	\left(g_{\mu\nu} - \frac{p_\mu p_\nu}{p^2}\right)
	G^{\mbox{\tiny LOR}}\left(\frac{p^2}{\mu^2}, g, \alpha \right) +
	\alpha \; \frac{p_\mu p_\nu}{p^2}\right).
\eeq
The renormalization group equation has the form
\beq
\label{eq:RenGroupLor}
\left( \mu^2 \frac{\partial}{\partial \mu^2} +
	\beta(g) \frac{\partial}{\partial g} +
	\delta(g) \frac{\partial}{\partial \alpha} +
	2 \gamma(g, \alpha) \right) G^{\mbox{\tiny LOR}}(\mu, \alpha, g) = 0,
\eeq
where 
\beqn
\beta(g) = \mu^2 \frac{\partial g}{\partial \mu^2},\ \ \  
\gamma(g) = \frac{\mu^2}{Z_A}\frac{d Z_A}{\mu^2}, \ \ \
\delta(g) = \mu^2 \frac{\partial \alpha}{\partial \mu^2}.
\eeqn
The gauge parameter $\alpha$ is considered as yet another 
coupling constant \cite{Logunov:1956}.

In the dimensional regularization approach, the renormalization group
functions $\beta(g)$ and $\gamma(g)$ are determined from the relations
\beqn
\beta(g)= {1\over 2}\left( g {d a_1 \over dg}\, - a_1 \right) ,\ \ \  
\gamma(g)= - {1 \over 2}\,g\, {dc_1 \over dg},
\eeqn
where $a_1$ and $c_1$ are the coefficients of $\ds {1\over \epsilon}$ 
in the $\ds {1\over \epsilon}$-expansion of the renormalization factors
$Z_A$ and $Z_g$:
\beqn
Z_g&=&1+{1\over g}\sum_{n=0}^\infty {a_n(g) \over \epsilon^n} \\ \nonumber
Z_A&=&1+\sum_{n=0}^\infty {c_n(g) \over \epsilon^n}\; .
\eeqn

In the one-loop approximation, the $\beta$ function is independent of
gauge and subtraction scheme. For the $SU(2)$ gauge field it is
given by 
\beq
\beta(g) = - bg^3 = -\, {11\over 3}\; {g^3 \over 16 \pi^2}.
\eeq 


In the Lorentz gauge, there are no contributions 
to the longitudinal component of the gluon polarization operator
and there is no counterterms to the gauge fixing term
in the Lagrangian. From this observation we obtain the relation between the 
renormalization factors, $Z_A^2 = Z_\alpha$, that determines
the dependence of the gauge parameter $\alpha$ on the normalization point.
In the case of the Landau gauge ($\alpha = 0$) one can
consider $\delta (g)=0$ (see \cite{Gross}); in the case 
$\alpha \neq 0$, the function $\delta (g)$ is not trivial. 

Here we solve the equation~(\ref{eq:RenGroupLor}) for
$\gamma = c g^2, \beta = - bg^3$.
First we note that, when $\gamma(g)=0$, any function
$F(z)$ with $\ds z={1\over g^2} + 2b\ln {p^2\over \mu^2}$ 
provides a solution of the equation~(\ref{eq:RenGroupLor}).
This function is specified by the initial (or boundary)
conditions. That is, if the sought-for function is given  
over a line transversal to characteristic curves of the 
equation then the function $F(z)$ can be determined and thus
the sought-for function is known over the $( g, {p^2\over \mu^2})$
plane.

A useful method of solution of the equations of the type
(\ref{eq:RenGroupLor}) with $\gamma(g) \neq 0$ 
is based on a treatment of the solutions of the respective 
auxiliary equation 
\beq\label{eq:RenGroupAux}
\left( - \frac{\partial}{\partial \lambda} + 
	\beta(g) \frac{\partial}{\partial g} + 
	\delta(g) \frac{\partial}{\partial \alpha} - 
	2 \gamma(g, \alpha)G^{\mbox{\tiny LOR}} \frac{\partial}{\partial G^{\mbox{\tiny LOR}}} \right) 
	V(\lambda, \alpha, g, G^{\mbox{\tiny LOR}}) = 0, 
\eeq 
where $\ds \lambda = \ln\frac{p^2}{\mu^2}$. Then the sought-for function 
$G^{\mbox{\tiny LOR}}(\lambda, g, \alpha)$ is determined from the condition
\beq\label{eq:SolutionII}
V(\lambda, \alpha, g, G^{\mbox{\tiny LOR}}) = 0, 
\eeq 
where $V(\lambda, \alpha, g, G^{\mbox{\tiny LOR}})$ is any solution 
of the auxiliary equation (\ref{eq:RenGroupAux}).
The solution of the auxiliary equation (\ref{eq:RenGroupAux}) 
is provided by any function which is constant along each characteristic curve. 
The characteristic curves are defined by the equation
\beq
\frac{d\lambda}{-1} = {dg \over \beta(g)} = {d\alpha \over \delta (g,\alpha)} =
-\, {dG^{\mbox{\tiny LOR}} \over 2\gamma(g,\alpha) G^{\mbox{\tiny LOR}}}.
\eeq
Integration of these equations gives
\beq\label{eq:CharLorIIa}
{1\over g^2} + {11\over 24\pi^2} \lambda = C_1,
\eeq
\beq\label{eq:CharLorIIb}
g^{-\;\left({13\over 11}\right)}\; {13- 3\alpha \over 3\alpha}  = C_2,
\eeq
\beq\label{eq:CharLorIIc}
{G^{\mbox{\tiny LOR}} \over \alpha} = C_3.
\eeq
Each of these equations defines three-dimensional hypersurface
and a characteristic curve is an intersection of these
hypersurfaces. Thus the solution of the equation (\ref{eq:RenGroupAux})
is given by 
\beq
V(\lambda, \alpha, g, G^{\mbox{\tiny LOR}}) = 
v \left({G^{\mbox{\tiny LOR}} \over \alpha} ,\;  \right. 
g^{-\;\left({13\over 11}\right)}\; {13- 3\alpha \over 3\alpha}, 
\left. {1\over g^2} + {11\over 24 \pi^2} \,\lambda \right) ,
\eeq
where $v(x,y,z)$ is an arbitrary sufficiently smooth function.

\begin{figure}[!h]
\centering
\includegraphics[angle=-90,scale=0.3]{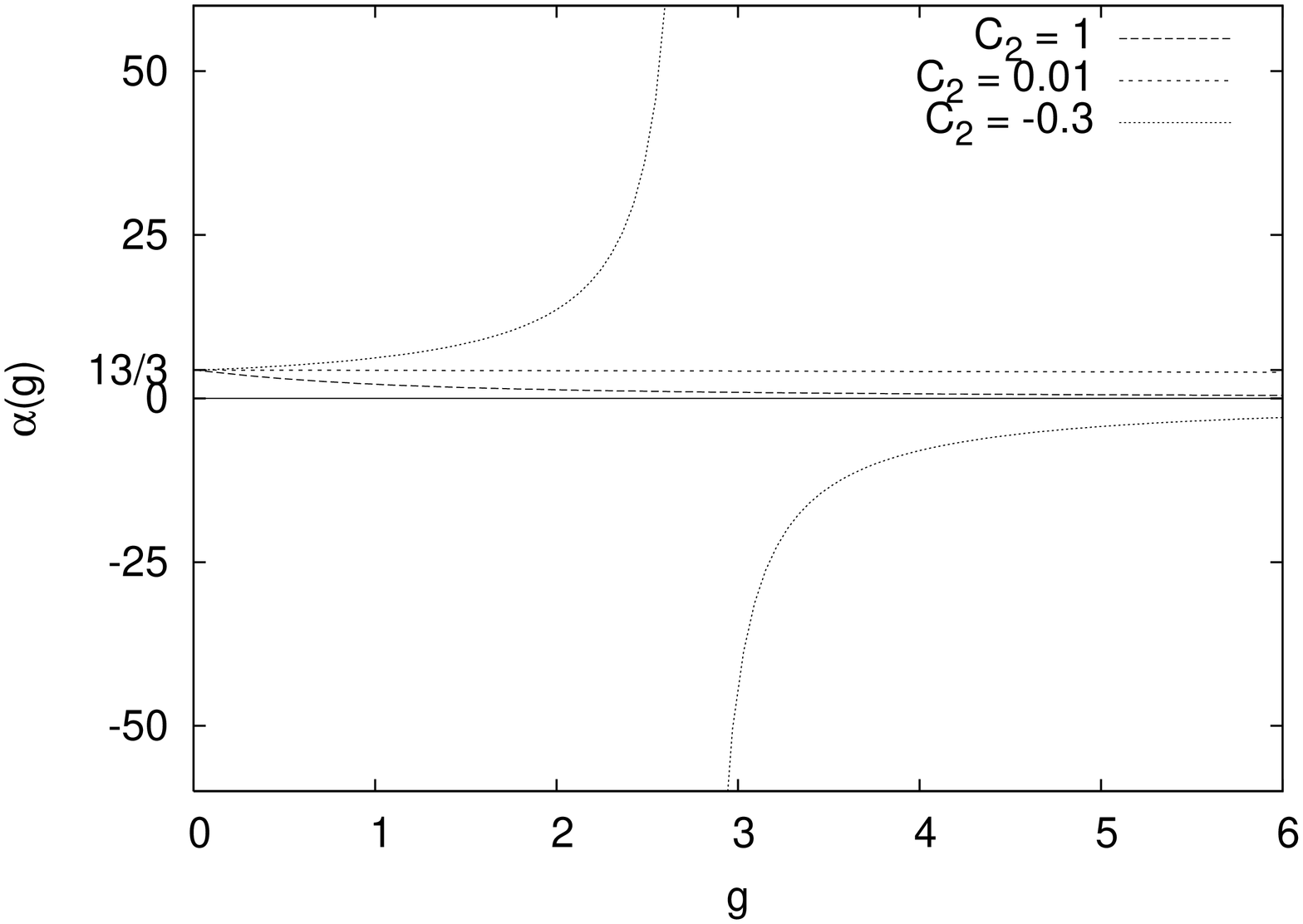}
\caption{Projection of characteristic curves onto the $(\alpha,g)$ plane.}
\label{fig:landau_char_curve}
\end{figure}

The condition (\ref{eq:SolutionII}) implies that 
\beq\label{eq:GenSolutionRGELor} 
G^{\mbox{\tiny LOR}}=\alpha \Phi\left(g^{\tiny -\;\left({13\over 11}\right)}\; {13- 3\alpha \over 3\alpha} \right., 
\left. {1\over g^2} + {11\over 24 \pi^2} \,\lambda \right) . 
\eeq 
Formula (\ref{eq:GenSolutionRGELor}), where $\Phi(y,z)$ is an
arbitrary function, represents a general solution of
the renormalization group equation in the Lorentz gauge.
A specific form of the function $\Phi(y,z)$ (a particular solution) 
is determined from the boundary conditions;
it is natural to fix them at the hyperplane $\lambda =0$. 
The behavior of the characteristic curves 
makes it possible to determine the asymptotic behavior 
of the propagator for $p^2 \to \infty $ and some fixed values of $g$ and $\alpha$
provided that its behavior for $g \to 0$, $\alpha \to 13/3$, and 
some fixed value of $\mu$ (say, $\mu^2 = p^2$) is known. 
In the latter domain, perturbation theory works
and thus the condition needed for the evaluation of the propagator at
large momenta is fulfilled. 
In the leading order of perturbation theory, 
$G=1$ and higher-order corrections are small.
For $\lambda =0$, we obtain the equation
\beq
\Phi(y,z)={1\over \alpha},
\eeq
which can be used for the determination of the dependence of $\Phi$ on
$y$ and $z$. On this hyperplane, 
$\ds y=g^{\tiny -\;\left({13\over 11}\right)}\; {13- 3\alpha \over 3\alpha}$ 
and $\ds z ={1\over g^2}$ and, therefore, 
$\ds \alpha = {13 \over 3 \left(1+yz^{\left(-\,{13 \over 22} \right)}\right)}\ $\ . 

Thus we arrive at
\beq
\Phi(y,z)={3\over 13}\, \left(1+yz^{\left(-\,{13 \over 22} \right)}\right),
\eeq

\beq\label{eq:LorGaugeProp8}
G^{\mbox{\tiny LOR}}\left(\frac{p^2}{\mu^2}, g, \alpha\right) =
{3\alpha \over 13}\,+\, \left( 1 -{3\alpha \over 13}\right)\,
\left(1+{11 g^2\over 24\pi^2}\,\ln\left( {p^2\over\mu^2} \right) \right)^{{\left(- \,{13\over 22} \right)}}. 
\eeq
Now we should take into account the scale dependence of the renormalized
quantities $\alpha$ and $g$. The expression for the running coupling has the form
\beq\label{eq:RunningCoupling0} 
g^2={24\pi^2 \over 11\; \ds \ln \left({\mu^2 \over \Lambda^2}\right)}\, . 
\eeq 
In the Lorentz gauge, the scale dependence of the gauge parameter
$\alpha$ (in other words, the dependence of the normalization point) 
is governed by the normalization condition as follows: 
the propagator at $p^2=\mu^2$ must have the form
\beq\label{eq:NormCondforProp}
G_{\mu\nu}^{AB}(p) = \frac{\delta^{AB}}{p^2} 
\left(\left(g_{\mu\nu}-\frac{p_\mu p_\nu}{p^2}\right)
+ \alpha_0\frac{p_\mu p_\nu}{p^2}\right)\, . 
\eeq
From the equations for the characteristic curves and the 
expression  (\ref{eq:RunningCoupling0}) for the running coupling
we obtain 
\beq 
\alpha = { 13 C \ln \left( {\mu^2 \over \Lambda^2} \right)^{(13/22)} \over
3 \left( 1+ C \ln \left( {\mu^2 \over \Lambda^2} \right)^{(13/22)} \right)},
\eeq
where the constant $C$ is determined from the normalization condition (\ref{eq:NormCondforProp}):
\beq
C={3\alpha_0 \over {\ds (13- 3 \alpha_0) \; \ln \left({p^2\over \Lambda^2}\right)^{{13\over 22}} } }\; .
\eeq
Thus the scale dependence of the gauge-fixing parameter is given by
\beq\label{eq:RunningGFParam}
\alpha(\mu)= {13\alpha_0 S \over 13 - 3\alpha_0\; (1-S)}, \ \ \ \ \ \ \ \ 
S=\left( \frac{\ln \left( \mu^2/\Lambda^2 \right)}
              {\ln \left(   p^2/\Lambda^2 \right)}
  \right)^{\frac{13}{22}}.
\eeq
The dependence of the propagator on the normalization point and
the momentum is obtained by substituting the expressions 
for the running gauge-fixing parameter (\ref{eq:RunningGFParam}) 
and running coupling (\ref{eq:RunningCoupling0}) into formula 
(\ref{eq:LorGaugeProp8}). After such substitution, we arrive at 
\beqn
G^{\mbox{\tiny LOR}}(S) = \frac{13 S}{13 - 3 \alpha_0 (1-S)}.
\eeqn

\subsection{The case of maximal Abelian gauge}

As contrasted from the case of Lorentz gauge,
the gluon propagator in the maximal Abelian gauge
cannot be represented in the form (\ref{eq:PropLor}); 
it can be parametrized by three functions as follows:
$G_{L}^{\off}, G_{T}^{\off}$, and $G^{(3)}$:
\beqn
G_{\mu\nu}^{33}(p) &=&
    \frac{1}{p^2}\left(
	\left(g_{\mu\nu}-\frac{p_\mu p_\nu}{p^2}\right) G^{(3)} +
	\hat\beta\; \frac{p_\mu p_\nu}{p^2} \right), \\
G_{\mu\nu}^{ab}(p) &=& \frac{\delta^{ab}}{p^2}\left(	
    \left(g_{\mu\nu}-\frac{p_\mu p_\nu}{p^2}\right) G^{\off}_T +
    \alpha\; \frac{p_\mu p_\nu}{p^2} G^{\off}_L\right). 
\eeqn

These functions are not independent because there should be
relations between them resulting from the Slavnov--Taylor identities.

The renormalization group equation for the one-particle irreducible 
Green's function has the form
\beqn\label{eq:RenGroup}
\left( \mu^2 \frac{\partial}{\partial \mu^2} +
	\beta(g) \frac{\partial}{\partial g} +
	\delta_\alpha(g) \frac{\partial}{\partial \alpha} +
	\delta_\beta(g) \frac{\partial}{\partial \hat \beta} +
	2 \gamma \right) G_2(\mu, \alpha, \hat \beta, g) = 0,
\eeqn
where $G_2$ is one of the functions $G_L^{\off}, G_T^{\off}$, 
or $G_T^{(3)}$ and the renormalization group functions are defined
by the equations
\beqn 
&& \beta(g) =\mu^2 {\partial g \over \partial \mu^2 }, \ \ \ 
\gamma(g)= \mu^2\; { d \ln Z_A \over d\mu^2}, \\ \nonumber
&& \delta_\alpha(g) =\mu^2 {\partial \alpha \over \partial \mu^2 }, \ \ \ 
\delta_\beta(g) =\mu^2 {\partial \hat \beta \over \partial \mu^2 }. \ \ \ 
\eeqn
In the one-loop approximation, we obtain
\beqn
\beta(g) &=& -b g^3, \\ \nonumber
\gamma^{(3)}(g) &=& c^{(3)} g^2 = -\;\frac{11}{3}\frac{g^2}{16 \pi^2},\\ \nonumber
\gamma^{o\!f\!f}(g) &=& c^{o\!f\!f} g^2 =
-\; \frac{(17-3\alpha-6\hat \beta )}{12} \frac{g^2}{16 \pi^2}, \\ \nonumber
\delta_\alpha(g) &=& -\; \frac{\left(3\alpha^2 - 4 \alpha + 9 \right)}{3}
\frac{g^2}{16 \pi^2}, \\ \nonumber
\delta_\beta(g)  &=& \frac{22}{3} \,\hat \beta \frac{g^2}{16 \pi^2 }.
\eeqn

The set of simultaneous equations defining the characteristic curves 
for the equation~(\ref{eq:RenGroup}) can be represented in a symmetric form,
\beqnn
\label{eq:CharMagSym}
	\frac{d\lambda}{-1} &=& \frac{d g}{\beta(g)} = 
\frac{d \alpha }{\delta_\alpha} = \frac{d \hat \beta }{\delta_\beta}=
\frac{d G_2}{-2 \gamma G_2}.
\eeqnn


Now we solve these equations for the Green's function of 
the "diagonal" gluon. Integration gives
\beqn
\label{eq:FirstInt1}
	I_1 &=& 2b\lambda + \frac{1}{g^2}, \\ \nonumber
	I_2 &=& g^2 \hat \beta,\\   \nonumber
	I_3 &=& \arctan \left(\frac{3\alpha-2}{\sqrt{23}}\right)-\frac{\sqrt{23}}{11}\ln g,\\ \nonumber
	I_4 &=& \frac{G_T^{(3)}}{\hat{\beta}},
\eeqn
\begin{figure}[!h]
\centering
\includegraphics[angle=-90,scale=0.3]{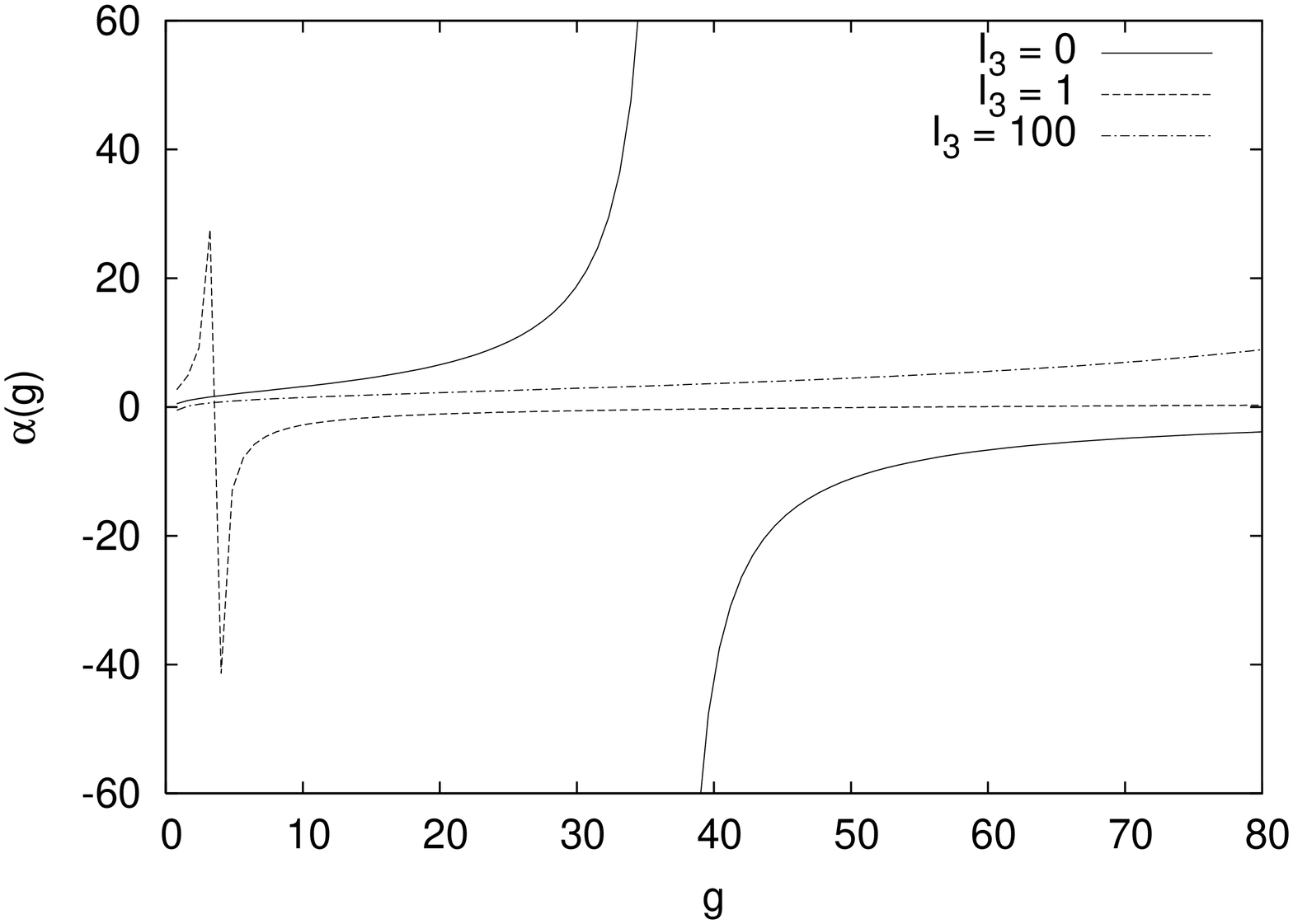}
\caption{Projection of characteristic curves onto the $(\alpha,g)$ plane.}
\label{fig:mag_char_curve}
\end{figure}
Thus the general solution of the equation~(\ref{eq:RenGroup}) has the form 
\beq
G_T^{(3)} = \hat{\beta}\; \Psi(I_3, I_2, I_1),
\eeq
where $\Psi(x,y,z)$ is an arbitrary function. The solution of the
equation~(\ref{eq:RenGroup}) for $G^{(3)}_L$ is the same. 
A particular solution is obtained by a determination of the specific
form of the function $\Psi(x,y,z)$ from the boundary condition. 
It is convenient to specify the boundary conditions on the 
plane $p^2=\mu^2$ ($\lambda = 1$) assuming that the perturbative
expansion for $G_T^{(3)}$ is valid at small $g$. 
In the tree approximation, $G_T^{(3)}=1$ and, therefore,
\beqn
\Psi(x,y,z) = \frac{1}{yz}.
\eeqn
From this formula it follows that 
\beqn
G_T^{(3)} = \frac{1}{1 + \frac{22}{3}\frac{g^2}{16 \pi^2}\ln\frac{p^2}{\mu^2}}.
\eeqn
Taking the scale dependence of $g^2$~(\ref{eq:RunningCoupling0}) into account yields 
\beq
\label{eq:mag_impr_prop}
G_T^{(3)} = \frac{\ln{\left(\mu^2/\Lambda^2\right)}}{\ln{\left(p^2/\Lambda^2\right)}}.
\eeq
The scale dependence of the parameter $\hat\beta$ can be determined from
the equations for the characteristic curves and formula~(\ref{eq:RunningCoupling0}),
the result is 
\beq
\hat\beta = \frac{11\,C_{\beta}}{24\,\pi^2} \ln\left(\frac{\mu^2}{\Lambda^2}\right).
\eeq
The parameter $C_{\beta}$ is determined from the normalization condition
$\hat\beta = \hat\beta_0$ at $p^2 = \mu^2$, LЁЁ-+
\beq
\hat\beta = \hat\beta_0\; \frac{\ln(\mu^2/\Lambda^2)}{\ln(p^2/\Lambda^2)}.
\eeq

The expression for the propagator for the "diagonal" gluon 
in the maximal Abelian gauge has a simpler form 
than the analogous expression in the Lorentz gauge. 
It can be accounted for by the remaining $U(1)$ 
symmetry in the maximal Abelian gauge, which implies
the relation $Z_g = Z_a^{-1/2}$ between the renormalization factors. 
Therefore, the functions $\beta(g)$ and $\gamma^{(3)}(g)$ 
are connected with each other by the formula $\beta(g) = g\, \gamma^{(3)}$.

Now we solve the equation~(\ref{eq:RenGroup}) 
for the transverse component of the Green's function
in the case of "off-diagonal" gluon. 
First integrals $I_1, I_2, I_3$ of the system~(\ref{eq:CharMagSym}) 
are the same as in the previous case. The fourth integral is given by 
\beq
I_4 = G_T^{\off} g^{ 15/22} e^{3\hat \beta \over 22}
        (3\alpha^2 - 4 \alpha + 9)^{-\,{1\over 4}}.
\eeq
Thus the solution of the equation~(\ref{eq:RenGroup}),
which is valid to both $G^{\off}_T$ and $G^{\off}_L$,
takes the form
\beq
G_T^{\off} = g^{\frac{15}{22}} e^{-\,\frac{3\hat\beta}{22}}(3\alpha^2 - 4 \alpha + 9)^{1/4}
\Phi(I_3,I_2,I_1),
\eeq
where $\Phi(x,y,z)$ is an arbitrary function, whose specific form is governed
by the boundary conditions. For the sake of convenience, 
we set the boundary conditions on the plane
$p^2 = \mu^2 (\lambda = 1)$.
It is assumed that, at small values of $g$, 
the perturbation theory expansion for the function $G_T^{\off}$ is valid.
In the tree approximation (that is, to zeroth order in $g$),
we obtain $G_T^{\off} = 1$ and
\beq
\Phi(x,y,z)=z^{-15/44} \exp ({3\over 22}\,yz) 
\left|{3\over 23}\; \cos\left(x-{\sqrt{23}\over 22}\ln z \right)\right|^{1\over 4}
\eeq
Therefore, 
\beq
G_T^{\off} \left({p^2\over \mu^2 }\, ,g,\alpha, \hat \beta \right) = 
\left({p^2\over \mu^2} \right)^{g^2\hat \beta /16\pi^2}
\xi^{15/44} \left|\cos\xi +{3\alpha -2 \over \sqrt{23}} \sin\xi\right|^{1/2},
\eeq
where
\beq
\xi={\sqrt{23}\over 22}\,\ln\zeta, \ \ \ 
\zeta=\left(1+{11 g^2 \over 24\pi^2}\,\ln \left({p^2\over \mu^2}\right)\right).
\eeq
The $\mu$ dependence of the parameters $\alpha$ and $\hat \beta$ 
is determined by the scale dependence of the running coupling
(we restrict out attention to the one-loop approximation)
and the values of the integration constants $I_2$ and $I_3$ in formulas
(\ref{eq:FirstInt1}). A specification of these values is equivalent to
specification of the normalization conditions, which, in the
case under consideration, is as follows:
the propagator at $p^2=\mu^2$ has the form 
\beq\label{eq:NormCondOff}
G_{\mu\nu}^{ab}=-i \delta^{ab}\left(\frac{1}{p^2}\left(g_{\mu\nu}-\frac{p_\mu p_\nu}{p^2}\right)
 + \hat\alpha_0 \frac{p_\mu p_\nu}{p^4}\right)\; .
\eeq
In view of the relation (\ref{eq:RunningCoupling0}), we obtain the equation
\beqn
\zeta={\ln (p^2/\Lambda^2) \over \ln (\mu^2/\Lambda^2)} 
\eeqn
giving the dependence of $\xi$ on $p^2$.  Though
it is a straightforward matter to derive the dependence of 
$\hat \beta$ on $p^2$, the $p^2$ dependence of $\alpha$ is derived in 
a more intricate way. To do this, we consider the factor 
$\ds  \left|\cos\xi +{3\alpha -2 \over \sqrt{23}} \sin\xi\right|^{1/2}$ 
and introduce the angle $\ds \varphi=\arctan {3\alpha -2 \over \sqrt{23}}$.
Then we obtain 
\beq
\left|\cos\xi +{3\alpha -2 \over \sqrt{23}} \sin\xi\right| =
\left|{\cos (\xi-\varphi)  \over \cos\varphi}\right| .
\eeq

From formula (\ref{eq:FirstInt1}) it follows that 
$\ds \varphi=I_3 + {\sqrt{23} \over 22} \ln g^2$ ($I_3$ is the 
integration constant independent of $\mu$). Taking the 
scale dependence of the coupling constant into account, 
we arrive at
\beq
\varphi(\mu^2)=I_3 - {\sqrt{23} \over 22} \ln \left( {11\over 24\pi^2} \ln (\mu^2/\Lambda^2) \right),
\eeq
where the constant $I_3$ is determined from the normalization condition at $p^2=\mu^2$:
\beq
\varphi(p^2)=\varphi_0=I_3 - {\sqrt{23} \over 22} \ln \left( {11\over 24\pi^2} \ln (p^2/\Lambda^2) \right).
\eeq
Thus we obtain 
\beq
\varphi(\mu^2)=\varphi_0+{\sqrt{23} \over 22} 
\ln \left( {\ln (p^2/\Lambda^2) \over \ln (\mu^2/\Lambda^2)}\right)
\eeq
and $\xi - \varphi = - \varphi_0$. 
From there formulas we derive the asymptotic behavior of the
"off-diagonal" propagator:
\beq \label{eq:UltimateResult}
G_T^{\off}=\left({p^2\over\mu^2}\right)^{- 3\hat \beta_0 \over 22\ln(p^2/\Lambda^2)}
\left({\ln (p^2/ \Lambda^2) \over \ln (\mu^2/ \Lambda^2)}\right)^{-\,{15\over 44}}
\left| {\cos\left( {\sqrt{23} \over 22}\ln\left({\ln(p^2/\Lambda^2)\over\ln (\mu^2/\Lambda^2)} \right) -\varphi_0 \right) \over \cos\varphi_0} \right|^{-\,1/2} 
\eeq 
The parameter $\varphi_0$ in this formula is determined by the relation 
$\ds \varphi_0 = \arctan {3\alpha_0 -2 \over\sqrt{23} }$,
where $\alpha_0$ is the gauge-fixing parameter at $p^2=\mu^2$ 
(see (\ref{eq:NormCondOff})).

\section{Discussion and conclusions}

The obtained expression (\ref{eq:UltimateResult}) is singular at
the values of momentum given by
\beqn
m(k) = \Lambda \left(\frac{\mu}{\Lambda}\right)^
{\exp\left\{\frac{22}{\sqrt{23}}\left(\frac{\pi}{2}+\varphi_0 + \pi k \right)\right\}},
\eeqn
where $k$ is an integer. 
An infinite number of singular 
points emerges both when 
$k \rightarrow -\infty\ \ \ (m(k) \rightarrow \Lambda)$ and when
$  k \rightarrow \infty\ \ \ (m(k) \rightarrow \infty)$.
In the neighborhood of a singular point,
the propagator behaves as $\ds G_T^{\off}\sim{1\over \sqrt{p^2 - m^2(k)}}$.
Neither such behavior nor the existence
of the singularities of a propagator in the Euclidean domain
has a simple physical interpretation. 
Such points are probably artifacts of the 
normalization condition used here, 
one-loop approximation, and the maximal Abelian
gauge in the continuum limit.
It should be noted that a mathematically rigorous
(say, as in  \cite{Zavialov}) consideration of the question of whether
this normalization condition is admissible can be
the subject of a separate study.
It should be emphasized that the maximal Abelian gauge
in the continuum limit, that is, the condition
\beq
(\partial_\mu + ig A^3_\mu) (A^1_\mu + iA^2_\mu)=0
\eeq
provides a relation between the fields of the order $A_\mu \sim 1$
with the fields of the order $\ds A_\mu \sim {1\over g}$,
which may cause difficulties in computations
in perturbation theory. 
The emergence of the terms $\ds {1\over \alpha}$
in the interaction Lagrangian is only one of manifestations 
of this relation. Such terms emerge in the gauge-fixing
part of the Lagrangian in a quite natural way.

For this reason, the maximal Abelian gauge is
essentially nonperturbative even in the domain of the small values of 
coupling constant. This conclusion is consistent with the
renormalization-group analysis of the propagator behavior
at the one-loop level. All the above implies that
the asymptotic behavior of the gluon propagator 
evaluated in this gauge on the lattice \cite{Morozov},
where nonperturbative contributions are naturally
taken into account, may substantially diverge
from the results of the renormalization-group improved 
computations in the framework of perturbation theory
performed in the present study. 

The propagator of the "diagonal" gluon evaluated on a lattice
\cite{Morozov} in the domain of large momenta can be adequately
parametrized by 
formula~(\ref{eq:mag_impr_prop}) with an appropriate choice of the
parameters $\Lambda, \mu$ and $\beta_0$.

The behavior of the "off-diagonal" propagator \cite{Morozov}
diverges substantially form our results. The reason is as follows:
\begin{enumerate}
\item There is no one by one correspondence between lattice MaG and 
perturbative MaG gauge;
\item Renormalization-group improved computations do not take into 
account nonperturbative contributions, which are naturally taken into 
account in the lattice calculations.
\end{enumerate}

However, a comprehensive comparison of the renormalization-group 
improved perturbative behavior of the propagator 
with the behavior obtained numerically in a lattice gauge theory
must include a consideration of the case when the gauge-fixing parameter 
$\alpha$ equals zero.
In this case, the quantization procedure in the
framework of the path-integral approach involves introducing 
an auxiliary field. A treatment of the gluon propagator
at $\alpha = 0$ may form the subject of a subsequent study.

\section*{Acknowledgments}

We are grateful to V.G.~Bornyakov, F.~V.~Gubarev, D.I.~Kazakov, M.I.~Polykarpov,   
and V.I.~Zakharov for helpful and stimulating discussions
and to A.A.~Slavnov for reading the manuscript and worth comments.
This study was supported in part by the Russian Foundation for Basic Research,
grant no 04-02-16079; S.M. also acknowledges the support of the Russian
Foundation for Basic Research, grant no. 03002-04016 and 03-02-16941.


\end{document}